# Deep learning-enabled multiplexed point-of-care sensor using a paper-based fluorescence vertical flow assay

*Artem Goncharov[1§], Hyou-Arm Joung[1§], Rajesh Ghosh[2], Gyeo-Re Han[1], Zachary S. Ballard[1], Quinn Maloney[1], Alexandra Bell[2], Chew Tin Zar Aung[3], Omai B. Garner[5], Dino Di Carlo[2,4]\* and Aydogan Ozcan[1,2,4]\**

[1]Electrical & Computer Engineering Department, [2]Bioengineering Department, [3] Microbiology, Immunology, & Molecular Genetics, [4]California NanoSystems Institute (CNSI), [5]Department of Pathology and Laboratory Medicine, University of California, Los Angeles, CA 90095 USA
[§]Contributed equally, *Corresponding Authors: ozcan@ucla.edu , dicarlo@ucla.edu

**Abstract:** We demonstrate multiplexed computational sensing with a point-of-care serodiagnosis assay to simultaneously quantify three biomarkers of acute cardiac injury. This point-of-care sensor includes a paper-based fluorescence vertical flow assay (fxVFA) processed by a low-cost mobile reader, which quantifies the target biomarkers through trained neural networks, all within <15 min of test time using 50 µL of serum sample per patient. This fxVFA platform is validated using human serum samples to quantify three cardiac biomarkers, i.e., myoglobin, creatine kinase-MB (CK-MB) and heart-type fatty acid binding protein (FABP), achieving less than 0.52 ng/mL limit-of-detection for all three biomarkers with minimal cross-reactivity. Biomarker concentration quantification using the fxVFA that is coupled to neural network-based inference is blindly tested using 46 individually activated cartridges, which showed a high correlation with the ground truth concentrations for all three biomarkers achieving > 0.9 linearity and < 15 % coefficient of variation. The competitive performance of this multiplexed computational fxVFA along with its inexpensive paper-based design and handheld footprint make it a promising point-of-care sensor platform that could expand access to diagnostics in resource-limited settings.

# Introduction

Centralized laboratory testing is a standard approach to the diagnostics of many common illnesses, including cardiovascular diseases, bacterial infections, and cancers. High accuracies and low sensitivities of such tests come at the expense of patient accessibility, long operating protocols and expensive equipment.[1] These factors impair applications of centralized diagnostics in low-resource, emergency and at-home settings due to economical or operational limitations.[2, 3] For example, cardiovascular diseases, while being the most common cause of mortality worldwide,[4] is disproportionally affecting (>70 %) people living in low- and middle-income countries with limited access to centralized medical care.[5] In such resource-limited settings, local medical facilities, if available, are often underfunded and may not have trained clinical laboratory technicians. Therefore, diagnostics tools for low-resource settings should be inexpensive and simple to operate.[6-7] Moreover, as emphasized by the COVID-19 pandemic[8-11] there is an increasing demand for more accessible diagnostic tests that can be operated by an individual at home to prevent the spread of viral diseases through timely diagnostics and management.[12-13] These ongoing challenges of global health invoke the need for affordable, user-friendly and yet robust point-of-care (POC) testing platforms for distributed and democratized diagnostics.

Paper-based diagnostic devices and sensors form an attractive platform for POC testing due to several attributes such as low-cost, low operating sample volumes, biocompatibility, ease of mass production and scalability to different chemistries and analytes.[13-15] In addition, all the fluid and sample transport through the paper materials occurs passively by capillary action, eliminating the need for fluidic pumps or mechanical components and enabling inexpensive designs that are easy-to-use and rapid.[16] Paper-based sensors have been implemented in different configurations[17] with the most popular being a lateral flow assay (LFA) format where the test solution flows laterally along the horizontally aligned paper layers.[18] In LFAs, the analytes of interest, usually protein antigens, are captured onto the paper surface through an affinity-based interaction using, for example, immunochemistry[18] and are further labeled by optical labels[19] for visual readout. For example, LFAs are widely used for pregnancy testing[20] and diagnostics of various diseases (e.g., HIV, influenza, COVID-19);[13, 21-23] however, relatively poor sensitivity and limited multiplexing of standard LFAs prevent their widespread use for diagnostics that require accurate quantitative measurements of a group of antigens and antigen panels at low concentrations.[24]

Standard LFAs rely on a colorimetric detection modality where the optical contrast results from the absorption of the incident light by gold nanoparticle labels (AuNPs);[25] sensitivity limitations in such LFAs are imposed by the absorption properties of AuNPs and limited density of AuNP labels within the testing region of the sensor [26]. Various fluorescent labels have also been employed in LFAs to replace AuNPs and achieve better sensitivity and quantitative analysis.[27-41] These fluorescence-based LFA platforms have been successfully applied for quantifying biomarkers from biological samples; however, they often rely on expensive readout hardware[27-29,34,36,38] and have low throughput with one or two analytes per sensor.[27-38] The limitation in biomarker multiplexing mainly comes from the *inline* geometry of LFAs, which constrains the flow length of the loaded samples and can lead to depletion of reagents and non-uniformity when flowing past adjacent test lines in series.[31]

Unlike LFAs, assays using the vertical flow of samples through stacked paper layers enable the arrangement of sensing regions in a 2D or 3D array and can achieve multiplexing with tens [42] or even hundreds [43] of independent testing channels represented by different affinity capture molecules. Vertical flow assays (VFAs) can be fabricated with non-trivial patterns of altering hydrophobic and hydrophilic regions that direct the liquid flow within the paper layers [44-46] and enable the incorporation of multiple sensing zones. Along with the developments in paper-based sensors, advances in computation and machine learning have recently emerged in the field of biomedical diagnostics with applications in histology,[47] digital pathology [48] and serodiagnosis of diseases [42, 49, 50], among others. Advanced machine learning techniques such as neural networks benefit from their universal approximation capability that can deal with noisy input data, making them a suitable candidate for quantifying POC sensors with multiplexed assays. Neural network-based inference and biomarker concentration measurements were previously applied for serodiagnosis using multiplexed vertical flow assays (xVFAs). [42, 49] These previous assays were based on a colorimetric detection modality (AuNPs) and used neural networks and multiple spots on the membrane to increase the dynamic range of the test for the quantitative detection of a *single* biomarker from human serum samples.[42]

Here we introduce a fluorescence-based VFA (fxVFA) for the multiplexed and simultaneous detection of three biomarkers from serum samples, accurately quantifying their concentrations using trained neural networks, all within <15 min of total testing time using 50 μL of serum sample per patient; see Figure 1. In our fxVFA design, we used conjugated polymer nanoparticles (CPNs) that have emerged in the past ~20 years as biocompatible and non-toxic fluorescent labels with tunable emission/excitation properties and minimally overlapping excitation and emission peaks.[51] The CPNs we used have 480 nm excitation and 610 nm emission peaks (Figure 2b), which helped us reduce the strong autofluorescence background from the paper substrate, which has been an issue, especially for quantum dot (QD) based assays due to their UV excitation.[33] Furthermore, owing to the presence of multiple absorption units and shared electronic structure within the polymer, the excitation energy transfer (FRET – Förster Resonance Energy Transfer) in CPNs happens across the whole backbone (through the molecular wire effect), catalyzing an amplified emission,[51] higher than for QDs (Figure 2c). CPNs are also more stable on porous paper layers with less photobleaching [52] and have larger sizes (e.g., >50 nm) compared to QDs leading to improved luminescence.[52]

We designed and validated the fxVFA platform for the simultaneous quantification of three biomarkers of cardiac injury, namely myoglobin, creatine kinase-MB (CK-MB), and heart-type fatty acid binding protein (FABP).[53] These markers are present in cardiac muscle and become elevated in the blood flow after acute myocardial infarction (AMI).[53] Typical reference ranges of these three markers in healthy individuals are 20-80 ng/mL for myoglobin,[54] 5-10 ng/mL for CK-MB[55] and 1-5 ng/mL for FABP,[56] making them challenging targets for colorimetric detection. Myoglobin and FABP are released after 1.5 hours from symptoms onset, have a peak concentration after 4-9 hours and might return to normal within 24 hours. Both proteins are also present in skeletal muscles and therefore positive detection (i.e. > 80 ng/mL for myoglobin and > 5 ng/mL for FABP) by itself should be used with caution for diagnosis of AMI since non-cardiac events can also cause their elevation.[53,54,56] CK-MB is an isoenzyme of creatine kinase present in the heart muscle and has optimal diagnostic performance for AMI in the period of 6-26 hours; however, it also becomes elevated after other cardiac pathologies (e.g., congestive cardiac failure), severe muscle damage (e.g., skeletal muscle disorders) and intake of drugs.[53] Therefore, the diagnosis of AMI based on the elevation of these three cardiac markers alone is risky; however, repeated negative tests in the relevant time intervals can be used to rule out AMI.[57, 58]

While some multiplexed paper-based fluorescence assays have been reported in the literature, including assays for the quantification of cardiac markers, [28,34-36,38-41] they demonstrated sensing of at most two analytes [32, 34] or employed relatively expensive benchtop readout hardware [30,31,36,39,40]. Assays with inexpensive handheld readers presented relatively inferior quantitative performance on clinical samples [34] or were only validated on buffer-spiked, artificial samples.[37, 39, 45] Our work presents a fluorescence-based VFA design and demonstrates a triplex biomarker sensor using a paper-based fluorescence assay with sub-ng/mL sensitivity and accurate concentration quantification on serum samples. Our fxVFA neural network models were trained using 30 unique serum samples and 4 control samples with 91 fxVFA cartridges that were activated. We blindly validated the performance of the trained fxVFA platform on 16 additional serum samples (not used during the training) measured in repeats by 46 individually activated fxVFA cartridges. We also created an inexpensive mobile phone-based reader to infer the concentrations of the target analytes through the trained neural networks. fxVFA measured concentrations presented a good correlation with the ground truth measurements from standard ELISA tests, achieving >0.9 coefficient of determination ($R^2$) and <15% average coefficient of variation (CV) for all three biomarkers. Taken together, the competitive performance of the fxVFA on serum samples, along with its low-cost design and simple operation that takes <15 min of testing time per patient using 50 μL of serum sample per test, makes it an appealing POC sensor for diagnostics in resource-limited settings.

## Results and Discussion

*Design of fxVFA*

The fluorescence-based vertical flow assay (fxVFA) is a paper-based sensor that consists of successive paper layers confined within a 3D-printed cartridge (Figure S1). The vertical design allows for the incorporation of a 2D array of testing spots on the sensing membrane, enabling multiplexed sensing with up to 100 individual immunoreaction channels. [42, 49] The sensing membrane of the fxVFA contains 17 isolated immunoreaction spots that are functionalized with capture antibodies specific to myoglobin, CK-MB, and FABP, as well as positive and negative controls. The assay takes less than 15 min to run, and consists of three loading steps, including the injection of a 50 µL sample mixed with CPN-Antibody (CPN-Ab) conjugates in the second step (see Experimental Methods). After the assay, an image of the activated sensing membrane is captured with a handheld fluorescence reader integrated on a mobile phone. Fluorescent signals from all 17 spots are further processed by a neural network-based quantification algorithm, which infers the concentrations of the biomarkers from the input signals. This fxVFA was optimized and validated for quantitative detection of three biomarkers (i.e., myoglobin, CK-MB, FABP) from serum samples and the results of these optimizations and validation are reported in the following sections.

*Optical properties of the CPN-based fxVFA*

We first quantified the higher brightness and improved sensitivity provided by CPNs compared to colorimetric detection (AuNPs) and other fluorescent labels (QDs) using the fxVFA platform (Figure 2). To accomplish this initial validation, all the candidate nanoparticles (i.e., CPNs, QDs, AuNPs) were functionalized with streptavidin (see the Experimental Methods section) and the sensing membrane of the assay was prepared with serially diluted concentrations of biotinylated BSA (b-BSA) that acts as the target sample analyte for this comparison assay (Figure 2d (i)). During the assay testing, streptavidin binds with b-BSA through the streptavidin-biotin interaction generating serially decreasing signals of labeled b-BSA as the concentration decreases from 50 µg/mL to 0 µg/mL (Figure 2d (ii)). After the assay activation for each cartridge, we captured its image to infer the lowest detectable concentration of b-BSA and determined the limit-of-detection (LOD, defined as the mean signal of the control + 3x standard deviation) for each optical label being compared (CPNs, QDs, AuNPs). When tested under the same illumination conditions, the LOD of this assay with CPN-based labels was over 3-fold lower than what was achieved using QDs (i.e., 0.06 µg/mL b-BSA for CPN and 0.2 µg/mL b-BSA for QDs). This resulted from the stronger autofluorescence of paper for the wavelengths used for the QD assay and higher fluorescence emission of CPNs (Figure 2d (ii)). Due to the size-dependent optical signal and flow properties of the CPNs through the paper layers, nanoparticles of different sizes yielded varying LODs. The optimal size of the CPNs for our fxVFA was determined to be ~70 nm based on the lowest LOD that was achieved (see Figure 2d (ii)).

*Sensitivity and specificity testing of fxVFA*

Before testing our CPN-based fxVFA for quantitative sensing of the three target biomarkers in human serum samples, we first validated its sensitivity and specificity in buffer-spiked samples. For this testing, we fabricated fxVFA cartridges, each with a sensing membrane composed of 17 immunoreaction spots, placed in a 2D array and separated with hydrophobic wax-printed barriers (see Experimental Methods). The spots were split between the testing conditions that specifically capture myoglobin, CK-MB, and FABP during the assay operation as well as a positive control condition that always stays at the same high-intensity level independent of the analyte concentration and a negative control condition that maintains a constant low-intensity level. The specificity of test spots comes from the immobilized capture antibodies spotted (i.e., against myoglobin, CK-MB or FABP). Two spots were used for each of these three biomarkers (i.e., repeats on the sensing membrane). The 11 remaining spots were split between positive (4 spots) and negative (7 spots) control conditions, where the positive control spots were functionalized with secondary anti-IgG antibodies and negative controls were functionalized with a spotting buffer (Figure 3a, Experimental Methods).

The operation of the triplex fxVFA takes <15 min per test and contains three sequential loading steps (see Experimental Methods). After the end of the assay, the bottom case of the sensor is inserted into a handheld mobile phone-based reader for image capture (Experimental Methods). To test the sensitivity of the fxVFA for the triplex panel, we used serially diluted samples and adjusted the concentration ranges of the spiked antigens to 0-300 ng/mL for myoglobin and 0-100 ng/mL for CK-MB and FABP. A larger dynamic range for myoglobin is selected due to the higher clinical reference range.[47] After the activation of these fxVFA cassettes (N=18, 6 samples x 3 repeats each) the captured fluorescence signals yielded a coefficient of determination of >0.99 with the spiked antigen concentrations for all three biomarkers; see Figure 3 (d-f). Based on these buffer spiking experiments, we calculated LODs as 0.52 ng/mL for myoglobin, 0.3 ng/mL for CK-MB, and 0.49 ng/mL for FABP (Figure 3 (d-f)).

In addition to the LOD experiments, we also tested the specificity of the fxVFA using spiked samples containing one or more biomarker proteins. We prepared three samples with the individually spiked antigens (one for myoglobin, one for CK-MB, and one for FABP) as well as one sample with all three antigens spiked together and one negative sample (Figure 3b). The spiking concentrations were 20 ng/mL for myoglobin, and 10 ng/mL for CK-MB and FABP. Only a minor increase in the fluorescence signal intensities was observed, corresponding to negligible cross-reactivity between non-matching antigen-antibody pairs as well as between CPN-Ab conjugates and the capture antibodies (Figure 3c).

LODs of the fxVFA directly depend on the repeatability of the assay and the level of the non-specific binding between the loaded samples and the immunoreaction spots on the sensing membrane. Therefore, before validating the fxVFA on spiked samples, we optimized its operation to account for these factors. To reduce the non-specific binding, we used an optimized spotting buffer (consisting of protein saver (1 mg/mL, PS) and Trehalose (1 %) in PBS) to functionalize the sensing membrane, which led to more than five-fold lower non-specific binding between the negative samples (i.e., 0 ng/mL) and the capture antibodies against CK-MB and FABP when compared to the assay with the control spotting buffer (PBS only) (Figure S2). This resulted in improved sensitivity owing to the higher specific signals from CK-MB and FABP in the positive samples (≥10 ng/mL concentration of each cardiac marker) (Figure S2). Moreover, the addition of PS and Trehalose into the running buffer along with the chicken ovalbumin (OVA, blocking reagent) yielded optimal inter-spot repeatability within the assay (Figure S3); therefore, we added these reagents into the optimal running buffer to reduce the sensor-to-sensor variations. Additionally, the repeatability of the assay was improved with a properly selected washing time. We optimized it to 12 min since it exhibited the lowest variation between the testing spots (Figure 1, Figure S4, see Experimental Methods).

*Training of fxVFA neural network-based biomarker quantification*

To accurately quantify our target biomarkers from serum samples and be resilient to various noise factors present in such samples (see e.g., Figure S5), we used deep learning and trained fully-connected neural networks to infer the cardiac marker concentrations from the fxVFA signals. fxVFA benefits from neural network-based information processing and inference to accurately quantify the target markers from more complex patient samples. For this task, the correct selection and optimization of the statistical learning model and its architecture are important to achieve reliable results. Before the final blinded testing of the fxVFA, we used a portion of the samples (validation set) to optimize the architecture of the neural network models and converged to three independently optimized fully-connected neural network models for the quantification of myoglobin ($DNN_{Myo}$), CK-MB ($DNN_{CK-MB}$) and FABP ($DNN_{FABP}$) (see Figure 4 and the Experimental Methods). We also optimized the input signals to these models by selecting an optimal subset of the 17 spots available on the paper membrane for each cardiac biomarker. This optimization process, referred to as *feature selection*, was implemented independently for each biomarker and incorporated two steps. In the first step, we optimized the number and the specific immunoreaction conditions from the three testing conditions (i.e., myoglobin, CK-MB, or FABP) along with three positive and three negative conditions per paper membrane. Different positive and negative conditions on the sensing membrane incorporated the same spotting immunochemistry and were differentiated based on their specific location within the sensing membrane. Location information includes: the inner part of the sensor (In; 1 positive and 2 negative spots), the outer part of the sensor (Out; 1 positive and 2 negative spots), and the corners (Cor; 2 positive and 2 negative spots); see Figure 5a. To form the neural networks' fluorescence input signals, we averaged alike spots within each condition. Using a 4-fold cross-validation approach, we selected an optimal set of conditions based on the biomarker concentration inference model with the lowest mean squared logarithmic error (MSLE) loss calculated between the predicted (i.e., neural network inferred) and the ground truth concentrations; see Figure 5 (b-d) for the optimal conditions selected for myoglobin, CK-MB and FABP.

After this condition selection step, next, we performed spot selection and ran this optimization step using an iterative backward feature selection process. In this process, we input the immunoreaction fluorescence signals of the validation data to the neural networks and eliminated spots one at a time at each iteration. fxVFA tends to keep the spot repeats benefiting from the averaged measurements; however, MSLE is reduced slightly as we removed the two negative control spots from $DNN_{CKMB}$ and one myoglobin test spot from $DNN_{FABP}$ (Figure 5 (e-h)). Based on this spot optimization, we selected the optimal sensing spots as 4 conditions with 7 spots for myoglobin (~41 % of the spots), 6 conditions with 9 spots for CK-MB (~53 % of the spots), and 3 conditions with 5 spots for FABP (~29 % of the spots) (see Figure 5(e-h), Figures S6-7 and Table S1). CK-MB has the highest number of spots (9 spots) in the optimal model, including 5 control spots (4 negative and 1 positive). This could be attributed to ~10x lower CK-MB concentration in most serum samples compared to myoglobin/FABP; its median concentration is 3.5 ng/mL versus 30 ng/mL for myoglobin and 17 ng/mL for FABP. Furthermore, CK-MB has ~5 times larger protein size (80 kDa vs. 18 kDa for myoglobin and 15 kDa for FABP), causing higher non-specific binding with positive and negative control spots. On the other hand, the optimal immunoreaction set for FABP had the lowest number of spots (5 spots) and did not contain any controls, probably due to its relatively small protein size (15 kDa) and lower non-specific binding with the positive and negative controls. The optimized set of spots for myoglobin incorporated 3 control spots, 2 FABP testing spots, and 2 specific myoglobin spots suggesting that myoglobin capture antibodies shared some cross-reactivity with the other 5 immunoreactions and that the neural network benefited from their combination for normalization and accurate concentration inference.

Predicted concentrations for the three analytes on the validation data set using this optimized fxVFA design and the associated neural network models are shown in Figure 4 and Figure 5 (i-k). This validation data set contains 20 unique serum samples, which were tested with $N_{val}$ = 57 fxVFA cartridges that we fabricated for this optimization. These tests are entirely separate from the blind testing samples and the corresponding fxVFA cartridges that were activated, which will be reported in the next section.

*Blind testing of the trained fxVFA*

The final neural network models were trained on the entire training data set comprising 30 unique serum samples and 4 negative control samples, totaling 91 activated fxVFA cartridges (see the Experimental Methods). The models were then tested on the previously unseen blind testing set comprising 16 additional serum samples (never seen before) and 46 individually activated fxVFA cartridges (see Figure 6 (b, d, f) and the Experimental Methods). For this blind testing phase, we used computationally optimized models with the optimal architectures and combinations of immunoreactions identified from the feature selection process (Figure 6 (a, c, e)). Ground truth concentrations for all the serum samples were quantified by ELISA, yielding for the test samples the corresponding concentration ranges of 8.6-75 ng/mL for myoglobin, 1.1-23.4 ng/mL for CK-MB and 1.3-45.7 ng/mL for FABP. Concentrations predicted by the fxVFA were highly correlated with the ground truth measurements across the entire clinical range with coefficients of determination (see Experimental Methods) of 0.92, 0.93, and 0.95 for myoglobin, CK-MB, and FABP, respectively (Figure 6 (b, d, f)). We also tested the repeatability of the fxVFA cartridges by running three repeats of each serum sample, and the average coefficients of variation (see Experimental Methods) between triplicate measurements were found to be 12.4 % for myoglobin, 12.6 % for CK-MB, and 12.5 % for FABP (Figure 6 (b, d, f)).

Importantly, the neural network models ($DNN_{Myo}$, $DNN_{CK-MB}$ and $DNN_{FABP}$) optimized through the feature selection process outperformed standard machine learning approaches, including multi-variable linear regression models as well as neural network models with non-optimal spot configurations (see Figure S8 and Tables S2-S4). During the feature selection process, our neural network models converged to an optimal spot outline that is not intuitive to a human operator and included a subset of the control spots as well as spots from the cross-reactive immunoreactions. Through this optimized design, deep learning models provided a statistically robust normalization that incorporates the reaction kinetics of different antigen/antibody pairs and the vertical flow properties of the injected liquids through the fxVFA cartridge design.

## Conclusions

In summary, we developed a fluorescence-based vertical flow assay (fxVFA) to simultaneously quantify myoglobin, CK-MB, and FABP in human serum samples. Our fxVFA has several advantages over other fluorescent paper-based assays. First, fxVFA benefits from the increased brightness of CPNs, resulting in a superior sensitivity compared to the assays with other fluorescent labels. [34, 38, 40] Additionally, unlike many existing fluorescent assays that require expensive benchtop readers for multiplexed sensing from biological samples, [27, 35, 39] our fxVFA uses a mobile phone-based handheld reader with competitive quantification of three biomarkers in human serum samples. Furthermore, the fxVFA platform is based on trained neural networks that can learn non-trivial relationships between the multiplexed fluorescent signals and the analyte concentrations, despite the presence of background noise in each test. Moreover, fxVFA has a user-friendly operation with three injection steps done through a single loading inlet, making it easy to implement even by a minimally trained technician using a custom operation kit. Finally, the assay uses 50 µL of serum sample per patient and takes under 15 min to complete, which is on the same scale as e.g., COVID-19 rapid antigen tests that take 15-30 min. Taken together, its competitive quantitative performance for multiplexed sensing of biomarkers from serum samples along with its easy and rapid operation, make our fxVFA a promising sensor for serodiagnosis in resource-limited and point-of-care settings.

## Experimental Methods Section

### fxVFA fabrication

#### Sensing membrane

The sensing membrane was prepared from a nitrocellulose membrane (Sartorius, USA) with a 2D pattern of 17 spots outlined by wax barriers printed by a wax printer (Xerox, Phase 8560). After the wax printing, the sensing membrane was incubated at 125 °C for 45 seconds in a forced-air dry oven (The Lab Depot, USA) to melt the wax and form the hydrophobic barriers. Spotting solutions (0.75 µL) were dispensed on each sensing spot according to the spotting map (Figure 3a) and dried for 10 min at 37 °C in a forced-air dry oven. All the spotting solutions were prepared with an optimized spotting buffer (see Figure S2). Test spots specific to myoglobin, CK-MB, and FABP were prepared as a mixture of the corresponding capture antibodies (2 mg/mL, Hytest, Fitzgerald) and the same volume of 2× spotting buffer. For the positive control spots, anti-mouse IgG (SouthernBiotech, USA) was diluted in the spotting buffer to 50 µg/mL, and for the negative control spots spotting solution was the spotting buffer only.

#### fxVFA cartridge preparation and assembly

The *fxVFA* cartridge was composed of top and bottom cases assembled through a twisting mechanism. The cases were 3D printed using a Formlabs 3D printer. The top case contained vertically stacked paper layers assembled in the following order from top to bottom: sampling membrane, vertical flow diffuser, 1$^{st}$ spreading layer, double-sided foam tape, interpad, 2$^{nd}$ spreading layer, and supporting layer (Figure S1). The bottom case was assembled by stacking five layers of absorption pad, double-sided foam tape, and sensing membrane (Figure S1). All the layers were cut with a laser cutter (Trotec, Speedy 100). The interpad, vertical flow diffuser and supporting layer were treated with BSA (1 % in PBS) solution to block hydrophobic interactions. Material information and cost for each layer are described in Table S5.

### Antibody-CPN conjugation

#### STA-CPN

CPNs (25 µL, CPN™ 610, Stream Bio Ltd., UK) were mixed with the reaction buffer (215 µL, 20 mM HEPES), and streptavidin (STA) (10 µL, 5 mg/mL). CPN mixtures were combined with PEG 8000 (5 µL, 5% v/v) and EDC (1 µL, 50 mg/mL) and incubated for 3 hours inside an orbital incubator (20 RPM). After incubation, glycine (2.5 µL, 10 mg/mL) was added and incubated for 30 min to block active sites. Next, after adding BSA (10 µL, 10% (w/v)) and DI water (200 µL), the mixture was centrifuged at 10,000g for 10 min using a microcentrifuge unit. After removing the solution, the CPN conjugates were re-suspended with incubation buffer (400 µL) and centrifuged at the same condition. After repeating the previous centrifugation step three times, the conjugates were re-suspended with the incubation buffer (250 µL). The brightness of the STA-conjugated CPNs was quantified using a well-plate reader.

#### Biotinylated antibody

Antibodies (100 µL, 2 mg/mL in PBS) were mixed with NHS-LC-Biotin (1 µL, 20 mM, Thermo Scientific) in DI water and then incubated for 30 min. The mixture was centrifuged at 14,000 g, 4 °C for 10 minutes using 10K centrifugal filter unit (10K MWCO, Thermo Scientific), then centrifuged again after discarding the solution and adding PBS (400 µL). After repeating the last step (i.e., the centrifugation with PBS) for at least 6 times, the biotinylated antibodies were centrifuged at 1000 g for 2 minutes and collected from the centrifugal filter (UFC501024, Millipore Sigma). The concentration of the synthesized antibody conjugates was measured by a well-plate reader (Synergy 2 Multi-Mode Microplate Reader, BioTek Instruments, Inc.).

*Antibody-CPN conjugation*

STA-CPN conjugates (60 µL, 1 mM) were mixed with biotinylated antibodies (20 µL, 1 mg/mL) and combined with BSA (200 µL, 1 mg/mL) in PBS. After two hours of incubation, Biotin-BSA (30 µL, mg/mL) and BSA (20 µL, 10% in PBS) were added and incubated for 30 min. After the incubation, BSA (300 µL, 1 % in PBS) was added and the mixture was centrifuged at 5000g for 10 min using 300K centrifugal filter unit (Vivaspin 20, 300,000 MWCO PES, Sartorius, USA). After discarding the filtered solution and adding BSA (300 µL 1% in PBS), the mixture was centrifuged again at the same condition. After repeating the previous step three times, PBS (300 µL) was added to restore CPN-Ab conjugates. The antibody-CPN conjugates were quantified with a well-plate reader.

*fxVFA operation*

Before each test, the top and bottom cases were assembled through a twisting mechanism. Then, the running buffer (200 µL) was loaded through the inlet at the top case. Next, the sample (50 µL, undiluted) was mixed with CPN-Antibody (CPN-Ab) conjugates (100 µL). After the absorption of the running buffer (< 30s) by paper layers, the sample with the conjugates mixture was loaded and allowed to absorb (~ 2 min). This step was followed by the addition of the washing buffer (400 µL) to wash away any non-specifically bound CPN-Ab conjugates and proteins. After loading the washing buffer (~12 min) the top and bottom cases were twisted apart and the bottom case was inserted into the handheld mobile phone-based reader for the image capture. The image of an activated sensing membrane was captured at a fixed imaging condition (0.5 s exposure, 50 ISO, manual focus to the sensing membrane) and saved in a raw format for subsequent processing.

*Handheld fluorescence reader*

The handheld reader was built from a mobile phone (LG G7 ThinQ) and a custom-designed 3D printed (Stratasys, USA) optical attachment (Figure 1b). The optical attachment consisted of a single emission channel located in front of the cell phone camera and four excitation units arranged in a circular shape around the emission channel (Figure 1b). Each excitation unit had 4 through-hole LEDs (CreeLED) with 480 nm peak emission wavelength coupled to the excitation filters (Edmund Optics). The LEDs were powered using a custom-designed constant current driver with 20 mA output current. The emission path consisted of an emission filter (Edmund Optics) and an external lens (Edmund Optics) to focus the sensing membrane onto the cell phone camera. A summary of all the optical components and their cost is presented in Table S6.

*Neural network-based processing*

All the 17 immunochemistry spots distributed on the sensing membrane were segmented from the captured mobile phone image using a custom automated segmentation code and the spot-wise fluorescent signal intensities were quantified as the mean pixel intensities:

$$(1) \quad I_m = \frac{1}{\Omega_p} \sum_{p \in \Omega_p} I_p,$$

$I_p$ refers to the per-pixel signal intensities and $\Omega_p$ defines the lateral bounds of the 2D spot mask for the $m^{th}$ immunoreaction spot. These 17 signals were then standardized according to the formula:

$$(2) \quad I_m^{st} = \frac{I_m - <I_m>}{\sigma(I_m)},$$

where $I_m$ is the intensity of the immunoreaction spot (m = 1 to 17), $<I_m>$ and $\sigma(I_m)$ are the mean and standard deviation of the $m^{th}$ immunoreaction spot calculated over the training set, respectively. These standardized signals served as inputs to the neural network models. The models were independently optimized for each of the three biomarkers through a 4-fold cross-validation on the training/validation set composed of 30 serum samples and 4 negative control samples with 91 individually activated fxVFA sensors/cartridges (see the *Serum sample testing* section below). The neural networks for each biomarker had fully connected layers optimized through a grid search of major neural network hyperparameters (e.g., number of hidden layers, number of units per layer, regularization, dropout, learning rate). As a result of this grid search, the optimal model for myoglobin had three hidden layers (256, 128, and 64 units, respectfully) and 0 dropout; for CK-MB – 2 hidden layers (256 and 64 units) with 0 dropout; for FABP – 3 hidden layers (256, 128 and 64 units) with 0.25 dropout. All three models used ReLU activation function and L2 regularizations for all the hidden layers. The optimal loss function for all models was the mean squared logarithmic error (MSLE) compiled with an Adam optimizer, a learning rate of 1e-3, and a batch size of 10. MSLE loss is defined as:

(3)
$$MSLE\ (y,\ y') = \frac{1}{N}\sum_{i=1}^{N}(\log(y_i + 1) - \log(y'_i + 1))^2,$$

where $y_i$ are the ground truth analyte concentrations, $y'_i$ are the predicted concentrations, and $N$ is the batch size (Figure 4). MSLE loss function was also used to compare the neural network models during the *feature selection* process (Figure 5, Results and Discussion) and the optimal model for each biomarker was selected based on the lowest MSLE score. Another metric that was used to evaluate and compare the neural network performance is the coefficient of determination ($R^2$, linearity or linear correlation), defined as:

(4)
$$R^2 = 1 - \frac{\sum_{i=1}^{N}(y_i - y'_i)^2}{\sum_{i=1}^{N}(y_i - <y>)^2},$$

where $<y>$ is the average of all ground truth concentrations. $R^2$ takes values between 0 and 1 and quantifies how well the predicted concentrations match the ground truth. In addition, we used the coefficient of variation (CV), which refers to the standard deviation between repeated measurements of the same serum sample divided by the mean of those measurements to evaluate the repeatability of the assay.

*Serum sample testing*

Remnant human serum samples were collected from UCLA Health in compliance with UCLA Institutional Review Board approved protocols (IRB#20-002084). Patient consent was waived since these are pre-existing remnant specimens collected independent of this research project. The samples were vortexed (~ 5s), centrifuged at 10,000g for 10 min and aliquoted (250 µL per aliquot) for storage at a -80 °C freezer. We tested 37 clinical and 17 spiked cTnI-free (Sigma) serum samples, each run in triplicate (x3) using our fxVFA, yielding 162 = 54x3 activated sensors. In addition to these 162 samples, we tested 4 PBS samples used as negative samples. Ground truth concentrations for all clinical and spiked serum samples were obtained from a parallel ELISA test. 2 out of 17 spiked serum samples and 6 out of 37 clinical serum samples had >100 ng/mL concentration of at least one of the target biomarkers which was beyond the linearity range of our sensor. Therefore, all three repeats of such samples were excluded from the final data set, yielding 46x3=138 fxVFAs activated with serum samples and 4 fxVFAs activated with PBS negative samples, which totals 142 activated fxVFA cartridges. Using 16 serum samples, we allocated 16x3=48 fxVFAs for blind testing of the final neural network models, leaving the remaining 94 tests for training and validation sets. Finally, we removed one fxVFA with abnormally high negative control signals and 4 fxVFAs with >20 % of CV between the positive control signals ending up with $N_{train}$ = 91 activated fxVFA sensors for training/validation and $N_{test}$ = 46 fxVFA sensors for blind testing.

*ELISA testing*

Capture antibodies (60 µL, 10 µg/mL in PBS) were dispensed to each well of the 96 well-plate and incubated for three hours. Then, BSA (300 µL, 1% in PBST) was added into each well and the well-plate was incubated overnight at 4 °C. Next, 5 times diluted samples were added (50 µL, 10 µL samples + 40 µL 1% BSA in PBST) and incubated for 2 hours which was followed by the addition of biotinylated detection antibodies (50 µL, 2 µg/mL) and incubation for 1 more hour. Further, STA-HRP solution (50 µL) was added to develop a color signal using TMB solution for 15 min. A UV well plate reader measured the optical density (OD) at 450 nm wavelength after adding the spot solution (50 µL, 0.2M sulfonic acid). We used a 150 rpm shaking incubator for all the incubation steps except for the BSA blocking (overnight incubation at 4 °C) and performed four times washing using PBST after the end of each incubation period.

*Acknowledgments*
The authors would like to acknowledge the US National Science Foundation (NSF) PATHS-UP Engineering Research Center (Grant #1648451) for funding. The authors also acknowledge Stream Bio (UK) for supplying fluorescent conjugated polymer nanoparticles (CPNs).

*Keywords*
Paper-based assays, vertical flow assays, multiplexed sensing, point-of-care sensing, conjugated polymer nanoparticles, cardiac biomarkers, deep learning, neural networks

**REFERENCES**
[1] S. S. Ehrmeyer, R. H. Laessig, *Clin. Chem. Lab. Med.* **2007**, *45*, 766;
[2] P. Yager, G. J. Domingo, J. Gerdes, *Annu. Rev. Biomed. Eng.* **2008**, *10*, 107;
[3] R. W. Peeling, D. Mabey, *Clinical microbiology and infection* **2010**, *16(8)*, 1062;
[4] M. Vaduganathan, G. A. Mensah, J. V. Turco, V. Fuster, G. A. Roth, *Journal of the American College of Cardiology* **2022**, *80(25)*, 2361;
[5] S. Yusuf, S. Islam, C. K. Chow, S. Rangarajan, G. Dagenais, R. Diaz, R. Gupta, R. Kelishadi, R. Iqbal, A. Avezum, A. Kruger, *The Lancet* **2011**, *378(9798)*, 1231;
[6] G. M. Whitesides, *Lab on a Chip* **2013**, *13(1)*, 11;
[7] V. Gubala, L. F. Harris, A. J. Ricco, M. X. Tan, D. E. Williams, *Analytical chemistry* **2012**, *84(2)*, 487;
[8] M. Chan-Yeung, R. H. Xu, *Respirology* **2003**, *8*, S9;
[9] G. Neumann, T. Noda, Y. Kawaoka, *Nature* **2009**, *459(7249)*, 931;
[10] S. T. Jacob, I. Crozier, W. A. Fischer, A. Hewlett, C. S. Kraft, M. A. D. L. Vega, M. J. Soka, V. Wahl, A. Griffiths, L. Bollinger, J. H. Kuhn, *Nature reviews Disease primers* **2020**, *6(1)*, 1;
[11] L. Yang, S. Liu, J. Liu, Z. Zhang, X. Wan, B. Huang, Y. Chen, Y. Zhang, *Signal transduction and targeted therapy* **2020**, *5(1)*, 1;
[12] P. Paliwal, S. Sargolzaei, S. K. Bhardwaj, V. Bhardwaj, C. Dixit, A. Kaushik, *Frontiers in Nanotechnology* **2020**, *2*, 571284;
[13] J. Budd, B. S. Miller, N. E. Weckman, D. Cherkaoui, D. Huang, A. T. Decruz, N. Fongwen, G. R. Han, M. Broto, C. S. Estcourt, J. Gibbs, D. Pillay, P. Sonnenberg, R. Meurant, M. R. Thomas, N. Keegan, M. M. Stevens, E. Nastouli, E. J. Topol, A. M. Johnson, M. Shahmanesh, A. Ozcan, J. J. Collins, M. F. Suarez, B. Rodriguez, R. W. Peeling, R. A. McKendry, *Nat Rev Bioeng* **2023**, 13–31;
[14] D. M. Cate, J. A. Adkins, J. Mettakoonpitak, C. S. Henry, *Analytical chemistry* **2015**, *87(1)*, 19;
[15] M. M. Gong, D. Sinton, *Chemical reviews* **2017**, *117(12)*, 8447;
[16] J. P. Rolland, D. A. Mourey, *MRS bulletin* **2013**, *38(4)*, 299;
[17] E. W. Nery, L. T. Kubota, *Analytical and bioanalytical chemistry* **2013**, *405(24)*, 7573;
[18] E. B. Bahadır, M. K. Sezgintürk, *TrAC Trends in Analytical Chemistry* **2016**, *82*, 286;
[19] B. B. Dzantiev, N. A. Byzova, A. E. Urusov, A. V. Zherdev, *TrAC Trends in Analytical Chemistry* **2014**, *55*, 81;
[20] T. Chard, *Human reproduction* **1992**, *7(5)*, 701;
[21] X. Fu, Z. Cheng, J. Yu, P. Choo, L. Chen, J. Choo, *Biosensors and Bioelectronics* **2016**, *78*, 530;
[22] A. T. Cruz, A. C. Cazacu, L. J. McBride, J. M. Greer, G. J. Demmler, *Annals of emergency medicine* **2006**, *47(3)*, 250;
[23] M. Sajid, A. N. Kawde, M. Daud, *Journal of Saudi Chemical Society* **2015**, *19(6)*, 689;
[24] A. Sena-Torralba, R. Álvarez-Diduk, C. Parolo, A. Piper, A. Merkoçi, *Chemical Reviews* **2022**, *122(18)*, 14881;
[25] M. Xu, B. R. Bunes, L. Zang, *ACS applied materials & interfaces* **2011**, *3(3)*, 642;
[26] G. R. Han, H. Ki, M. G. Kim, *ACS applied materials & interfaces* **2019**, *12(1)*, 1885;
[27] A. N. Berlina, N. A. Taranova, A. V. Zherdev, Y. Y. Vengerov, B. B. Dzantiev, *Analytical and bioanalytical chemistry* **2013**, *405(14)*, 4997;

[28] D. Lou, L. Fan, Y. Cui, Y. Zhu, N. Gu, Y. Zhang, *Anal. Chem.* **2018**, *90*, 6502;
[29] J. Hu, Y. Z. Jiang, L. L. Wu, Z. Wu, Y. Bi, G. Wong, X. Qiu, J. Chen, D. W. Pang, Z. L. Zhang, *Analytical chemistry* **2017**, *89(24)*, 13105;
[30] J. Liu, D. Ji, H. Meng, L. Zhang, J. Wang, Z. Huang, J. Chen, J. Li, Z. Li, *Sensors and Actuators B: Chemical* **2018**, *262*, 486;
[31] X. Guo, Y. Yuan, J. Liu, S. Fu, J. Zhang, Q. Mei, Y. Zhang, *Analytical Chemistry* **2021**, *93(5)*, 3010;
[32] X. Qi, Y. Huang, Z. Lin, L. Xu, H. Yu, *Nanoscale research letters* **2016**, *11(1)*, 1;
[33] H. He, B. Liu, S. Wen, J. Liao, G. Lin, J. Zhou, D. Jin, *Analytical chemistry* **2018**, *90(21)*, 12356;
[34] M. You, M. Lin, Y. Gong, S. Wang, A. Li, L. Ji, H. Zhao, K. Ling, T. Wen, Y. Huang, D. Gao, *ACS nano* **2017**, *11(6)*, 6261;
[35] X. Cheng, X. Pu, P. Jun, X. Zhu, D. Zhu, M. Chen, *International journal of nanomedicine* **2014**, *9*, 5619;
[36] Y. Cai, K. Kang, Q. Li, Y. Wang, X. He, *Molecules* **2018**, *23(5)*, 1102;
[37] M. Ren, H. Xu, X. Huang, M. Kuang, Y. Xiong, H. Xu, Y. Xu, H. Chen, A. Wang, *ACS applied materials & interfaces* **2014**, *6(16)*, 14215;
[38] M. D. Wilkins, B. L. Turner, K. R. Rivera, S. Menegatti, M. Daniele, *Sensing and bio-sensing research* **2018**, *21*, 46;
[39] J. Zou, X. Liu, X. Ren, L. Tan, C. Fu, Q. Wu, Z. Huang, X. Meng, *Nanoscale* **2021**, *13(16)*, 7844;.
[40] L. Huang, Y. Zhang, E. Su, Y. Liu, Y. Deng, L. Jin, Z. Chen, S. Li, Y. Zhao, N. He, *Nanoscale Advances* **2020**, *2(3)*, 1138;
[41] X. Guo, L. Zong, Y. Jiao, Y. Han, X. Zhang, J. Xu, L. Li, C. W. Zhang, Z. Liu, Q. Ju, J. Liu, *Analytical chemistry* **2019**, *91(14)*, 9300;
[42] Z. S. Ballard, H. A. Joung, A. Goncharov, J. Liang, K. Nugroho, D. Di Carlo, O. B. Garner, A. Ozcan, *NPJ digital medicine* **2020**, *3(1)*, 1;
[43] A. W. Martinez, S. T. Phillips, G. M. Whitesides, *Proceedings of the National Academy of Sciences* **2008**, *105(50)*, 19606;
[44] S. Mohammadi, M. Maeki, R. M. Mohamadi, A. Ishida, H. Tani, M. Tokeshi, *Analyst* **2015**, *140(19)*, 6493;
[45] Y. Jiao, C. Du, L. Zong, X. Guo, Y. Han, X. Zhang, L. Li, C. Zhang, Q. Ju, J. Liu, H. D. Yu, *Sensors and Actuators B: Chemical* **2020**, *306*, 127239;
[46] W. Dungchai, O. Chailapakul, C. S. Henry, *Analyst* **2011**, *136(1)*, 77;
[47] Y. Rivenson, H. Wang, Z. Wei, K. de Haan, Y. Zhang, Y. Wu, H. Günaydın, J. E. Zuckerman, T. Chong, A. E. Sisk, L. M. Westbrook, W. D. Wallace, A. Ozcan, *Nature biomedical engineering* **2019**, *3(6)*, 466;
[48] K. Bera, K. A. Schalper, D. L. Rimm, V. Velcheti, A. Madabhushi, *Nature reviews Clinical oncology* **2019**, *16(11)*, 703;
[49] H. A. Joung, Z. S. Ballard, J.Wu, D. K. Tseng, H. Teshome, L. Zhang, E. J. Horn, P. M. Arnaboldi, R. J. Dattwyler, O. B. Garner, D. Di Carlo, A. Ozcan, *ACS nano* **2019**, *14(1)*, 229;
[50] Y. Luo, H. A. Joung, S. Esparza, J. Rao, O. Garner, A. Ozcan, *Lab on a Chip* **2021**, *21(18)*, 3550.
[51] S. W. Thomas, G. D. Joly, T. M. Swager, *Chemical reviews* **2007**, *107(4)*, 1339;
[52] C. Zhu, L. Liu, Q. Yang, F. Lv, S. Wang, *Chemical reviews* **2012**, *112(8)*, 4687;
[53] H. A. Al-Hadi, K. A. Fox, *Sultan Qaboos University Medical Journal* **2009**, *9(3)*, 231;
[54] F. A. Van Nieuwenhoven, A. H. Kleine, K. W. H. Wodzig, W. T. Hermens, H. A. Kragten, J. G. Maessen, C. D. Punt, M. P. Van Dieijen, G. J. Van Der Vusse, J. F. Glatz, *Circulation* **1995**, *92(10)*, 2848;
[55] M. Van Blerk, V. Maes, L. Huyghens, M. P. Derde, R. Meert, F. K. Gorus, *Clinical chemistry* **1992**, *38(12)*, 2380;
[56] J. F. Glatz, G. J. Van der Vusse, J. G. Maessen, V. D. V. MP, W. T. Hermens, *Acta anaesthesiologica Scandinavica. Supplementum* **1992**, *111*, 292;
[57] A. P. Varki, D. S. Roby, H. Watts, J. Zatuchni, *American heart journal* **1978**, *96(5)*, 680;
[58] A. Colli, M. Josa, J. L. Pomar, C. A. Mestres, T. Gherli, *Cardiology* **2007**, *108(1)*, 4.

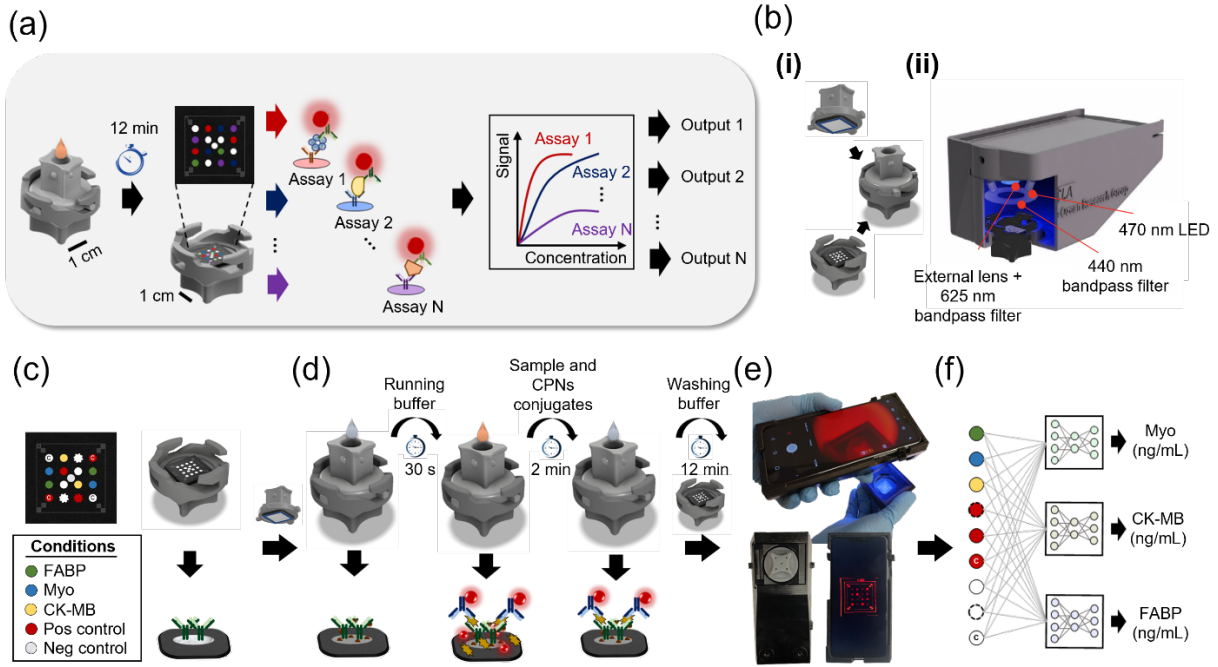

**Figure 1**. (a) Overview of the multiplexed paper-based fluorescence vertical flow assay (fxVFA). (b) fxVFA cartridge (i) and hand-held fluorescence reader (ii). (c) Outline of the fxVFA sensing membrane for the simultaneous sensing of myoglobin, CK-MB and FABP. (d) Operation of the fxVFA. The assay operation takes under 15 min and contains three sequential loading steps including the loading of 50 μL sample mixed with CPN-Ab conjugates at the second step. (e) Sensing membrane image capture with the hand-held fluorescence reader. (f) Neural network-based biomarker quantification using the fully-connected trained networks.

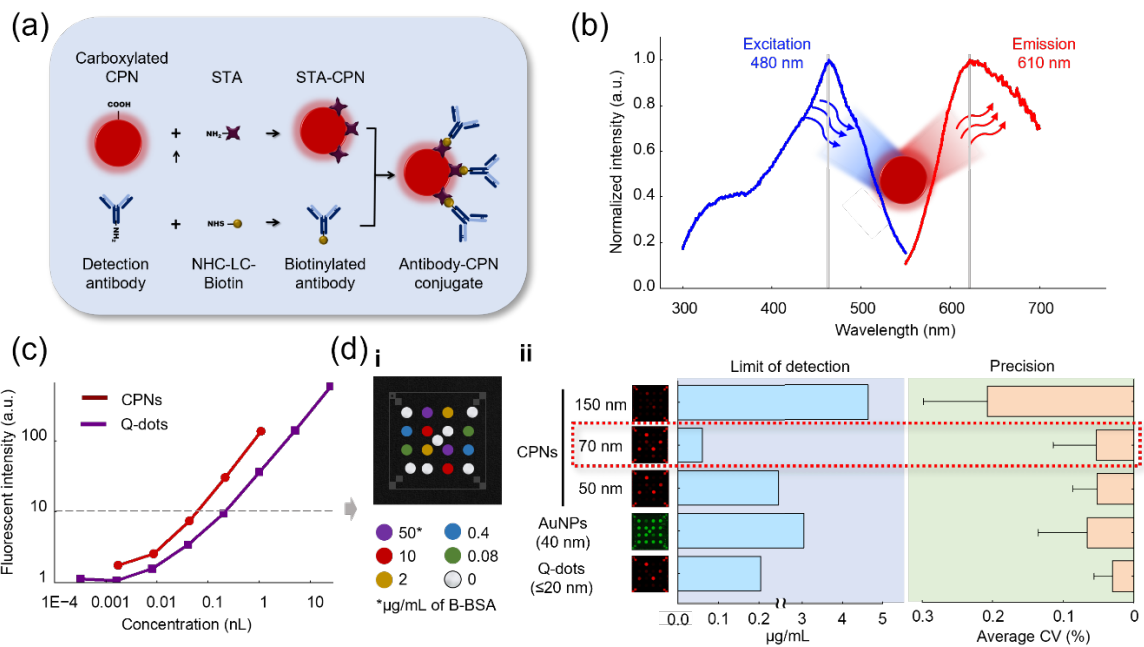

**Figure 2**. (a) Conjugation procedure between CPNs and antibodies. (b) Absorption and emission spectra of CPNs. (c) Brightness comparison between colloidal CPNs and Q-dots. (d) (i) Spotting map outline for the sensitivity evaluation of different labels at fxVFA; (ii) LODs of the fxVFA for the labels of interest (left) and the average CVs of the fluorescent and colorimetric signals from the spot repeats with the same b-BSA concentrations (right).

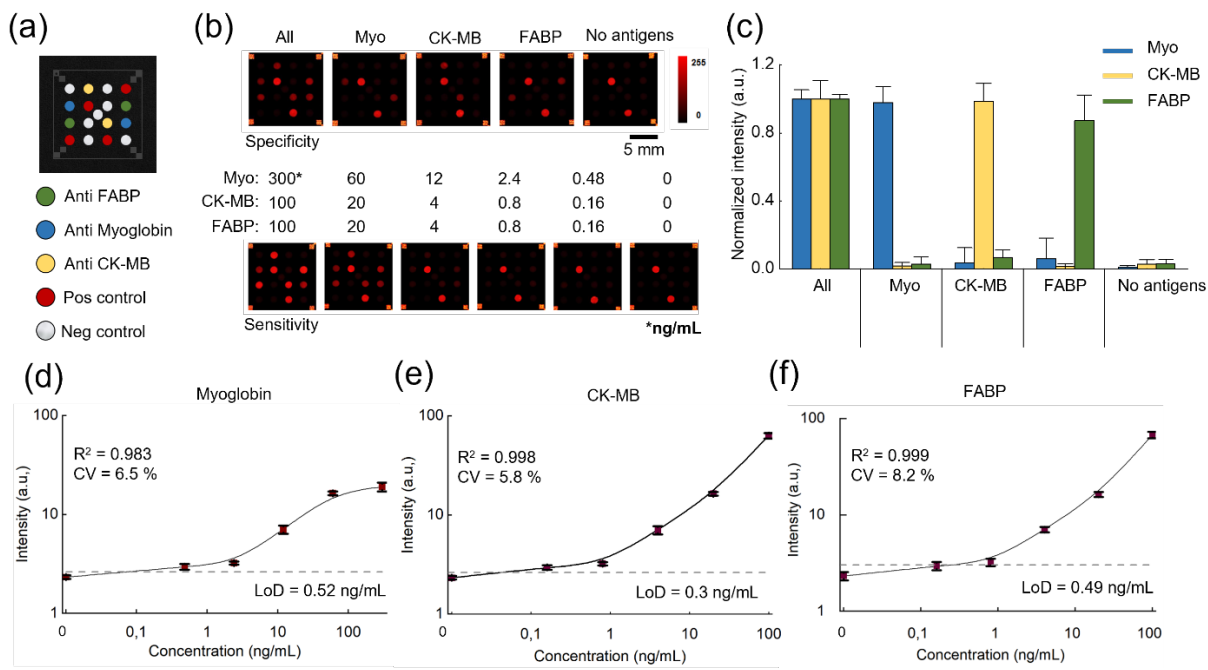

**Figure 3**. (a) Spotting map at the fxVFA membrane for the sensitivity and specificity testing. (b) Images of the activated sensing membranes for specificity (top) and sensitivity (bottom) tests. (c) Specificity results for the multiplexed detection of myoglobin, CK-MB and FABP at the fxVFA. Sensitivity results of the fxVFA for quantitative detection of (d) myoglobin, (e) CK-MB, and (f) FABP. All the samples used for sensitivity and specificity testing are buffer-spiked samples containing the biomarkers of interest.

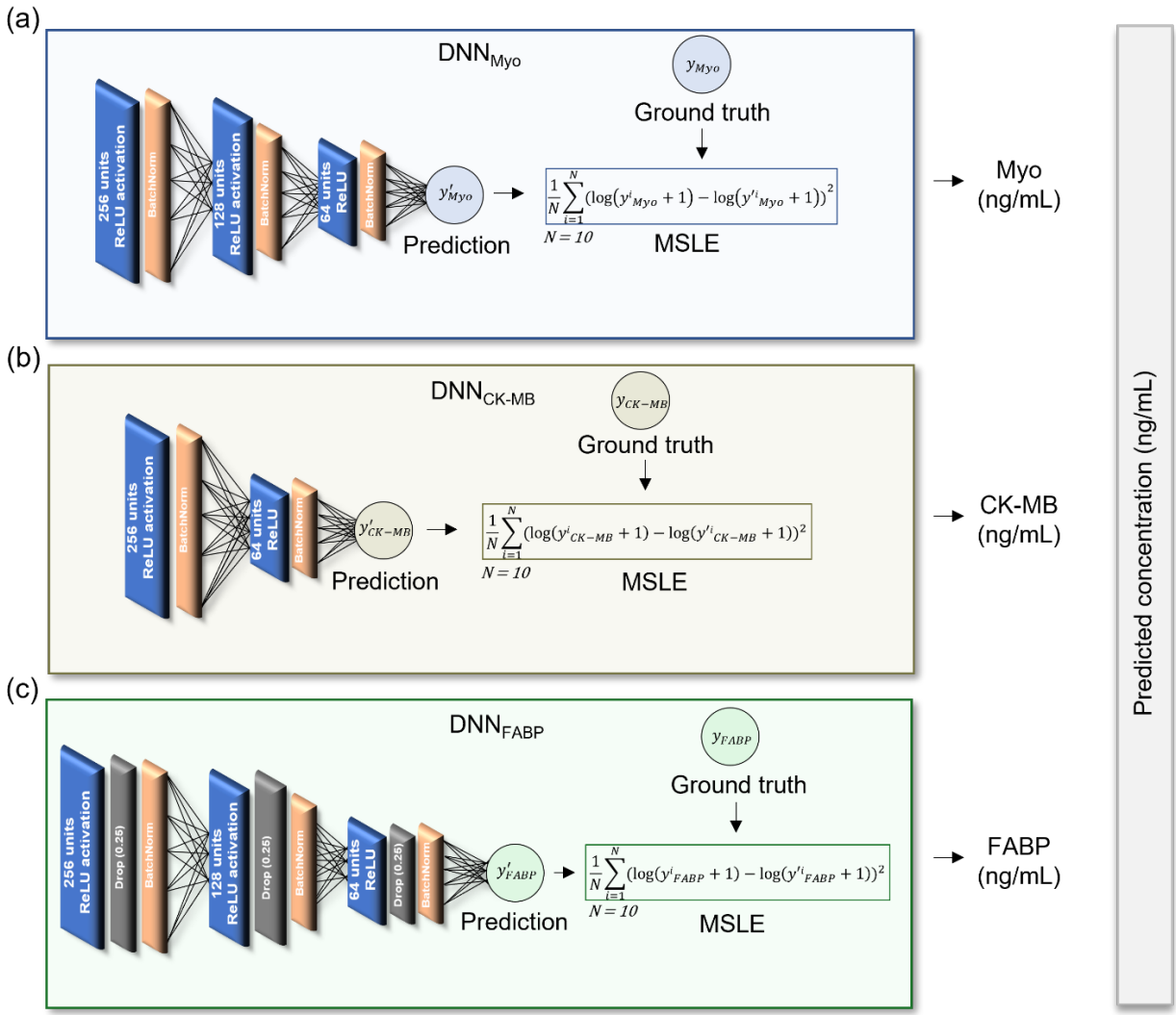

**Figure 4**. Fully-connected neural network architectures that are trained for the quantification of (a) myoglobin (DNN$_{Myo}$), (b) CK-MB (DNN$_{CK-MB}$), and (c) FABP (DNN$_{FABP}$).

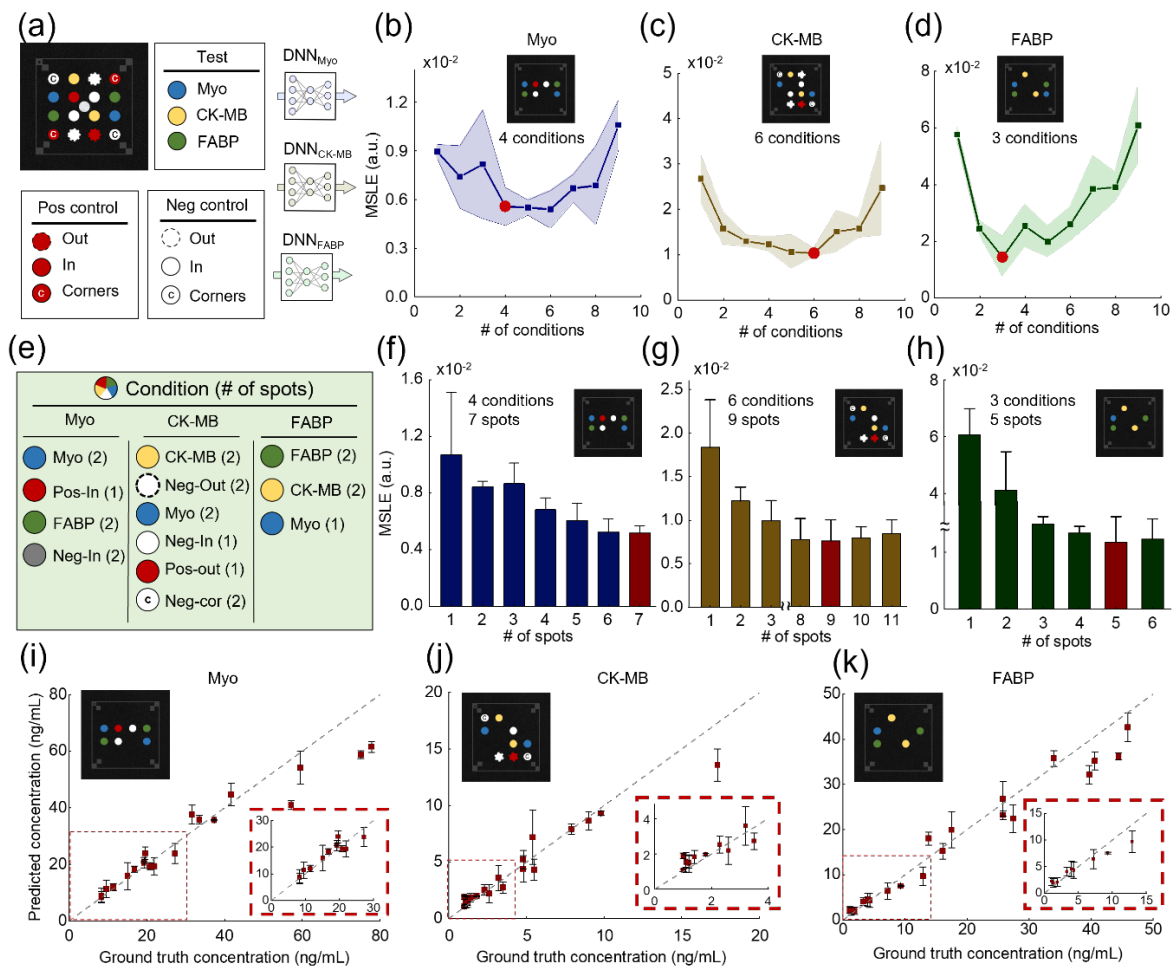

**Figure 5**. Computational optimization of the input signals to the neural network models through the *feature selection* process. (a) Sensing membrane outline for the feature selection process. (b-d) First step of the feature selection process – condition selection. Neural network models with different subsets of conditions are compared based on the mean squared logarithmic error (MSLE) loss from the 4-fold cross-validation on the validation data set ($N_{val}$ = 57). MSLE losses of selected models are plotted against the number of input conditions for (b) myoglobin, (c) CK-MB, and (d) FABP. Optimal set of conditions for each biomarker is selected based on the lowest MSLE loss from the cross-validation and is indicated by a solid red marker. Insets at each plot show the configurations of the sensing membranes with the optimal sets of conditions. (e) Outline of the optimal conditions/spots for each biomarker identified from the feature selection process. (f-h) Second step of the feature selection process – spot selection. Spots are eliminated iteratively through a backward feature selection process. MSLE losses for the models at each iteration are plotted against the number of input spots for (f) myoglobin, (g) CK-MB, and (h) FABP. Optimal spot configuration is selected based on the model with the lowest MSLE loss, highlighted with red color. Insets at each plot show the configurations of the sensing membranes with the optimal set of sensing spots. (i-k) Ground truth concentrations (x-axis) against the predicted concentrations (y-axis) resulting from the optimized neural network models with the optimal input spots and conditions (i.e. $DNN_{Myo}$, $DNN_{CK-MB}$ and $DNN_{FABP}$) tested on the validation data set ($N_{val}$ = 57) for (i) myoglobin, (j) CK-MB, and (k) FABP.

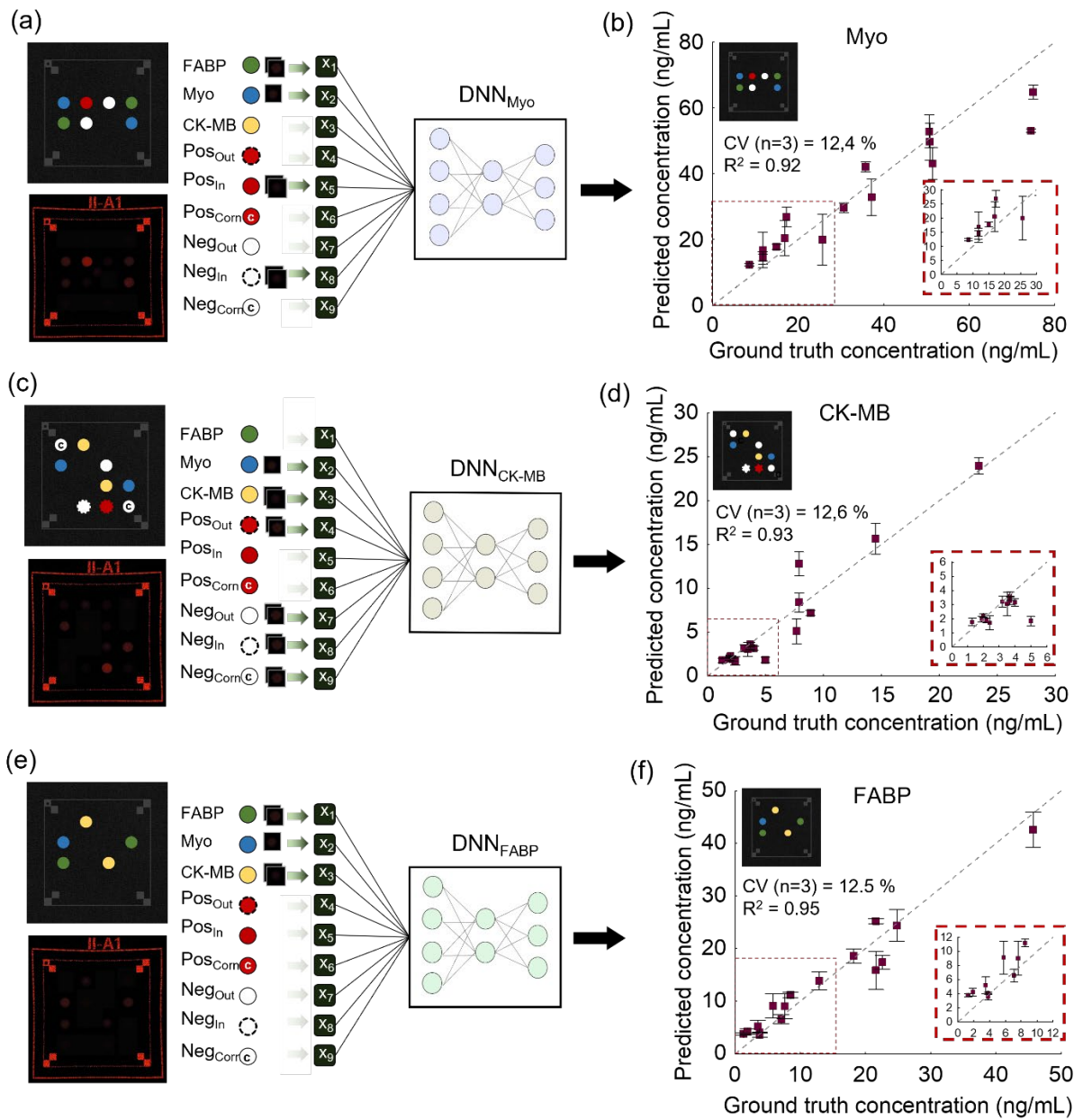

**Figure 6**. Computationally optimized neural network models with the optimal architectures and combinations of immunoreactions identified from the feature selection process for the quantification of (a) myoglobin ($DNN_{Myo}$), (c) CK-MB ($DNN_{CK-MB}$), and (e) FABP ($DNN_{FABP}$). fxVFA predicted concentrations (y-axis) are plotted against the ground truth concentrations (x-axis) for the blind testing data set of serum samples ($N_{test}$ = 46) for (b) myoglobin, (d) CK-MB, and (f) FABP. The dashed lines indicate y = x.

17